\documentclass[aps,prl,twocolumn,superscriptaddress,showpacs]{revtex4}
\usepackage{graphicx}
\usepackage{mathrsfs}
\usepackage{bm}
\usepackage{amsmath}
\usepackage{dcolumn}
\usepackage{epstopdf}
\usepackage{dsfont}
\usepackage{amssymb}
\usepackage{tabularx}
\usepackage{array}
\usepackage{float}
\usepackage{colordvi}

\begin{document}
\title{In-Plane Magnetization Induced Quantum Anomalous Hall Effect in Atomic Crystals of Group-V Elements}
\author{Peichen Zhong$^\ddagger$}
\affiliation{ICQD, Hefei National Laboratory for Physical Sciences at Microscale, and Synergetic Innovation Center of Quantum Information and Quantum Physics, University of Science and Technology of China, Hefei, Anhui 230026, China.}
\affiliation{CAS Key Laboratory of Strongly-Coupled Quantum Matter Physics, and Department of Physics, University of Science and Technology of China, Hefei, Anhui 230026, China.}
\author{Yafei Ren$^\ddagger$}
\affiliation{ICQD, Hefei National Laboratory for Physical Sciences at Microscale, and Synergetic Innovation Center of Quantum Information and Quantum Physics, University of Science and Technology of China, Hefei, Anhui 230026, China.}
\affiliation{CAS Key Laboratory of Strongly-Coupled Quantum Matter Physics, and Department of Physics, University of Science and Technology of China, Hefei, Anhui 230026, China.}
\author{Yulei Han}
\affiliation{ICQD, Hefei National Laboratory for Physical Sciences at Microscale, and Synergetic Innovation Center of Quantum Information and Quantum Physics, University of Science and Technology of China, Hefei, Anhui 230026, China.}
\affiliation{CAS Key Laboratory of Strongly-Coupled Quantum Matter Physics, and Department of Physics, University of Science and Technology of China, Hefei, Anhui 230026, China.}
\author{Liyuan Zhang}
\email[Corresponding author:~~]{zhangly@sustc.edu.cn}
\affiliation{Department of Physics, South University of Science and Technology of China, Shenzhen, Guangdong 518055, China.}
\author{Zhenhua Qiao}
\email[Corresponding author:~~]{qiao@ustc.edu.cn}
\affiliation{ICQD, Hefei National Laboratory for Physical Sciences at Microscale, and Synergetic Innovation Center of Quantum Information and Quantum Physics, University of Science and Technology of China, Hefei, Anhui 230026, China.}
\affiliation{CAS Key Laboratory of Strongly-Coupled Quantum Matter Physics, and Department of Physics, University of Science and Technology of China, Hefei, Anhui 230026, China.}

\begin{abstract}
  We theoretically demonstrate that the in-plane magnetization induced quantum anomalous Hall effect (QAHE) can be realized in atomic crystal layers of group-V elements with buckled honeycomb lattice. We first construct a general tight-binding Hamiltonian with $sp^3$ orbitals via Slater-Koster two-center approximation, and then numerically show that for weak and strong spin-orbit couplings the systems harbor QAHEs with Chern numbers of $\mathcal{C}=\pm1$ and $\pm2$ , respectively. For the $\mathcal{C}=\pm1$ phases, we find the critical phase-transition magnetization from a trivial insulator to QAHE can become extremely small by tuning the spin-orbit coupling strength. Although the resulting band gap is small, it can be remarkably enhanced by orders via tilting the magnetization slightly away from the in-plane orientation. For the $\mathcal{C}=\pm2$ phases, we find that the band gap is large enough for the room-temperature observation. Although the critical magnetization is relatively large, it can be effectively decreased by applying a strain. All these suggest that it is experimentally feasible to realize high-temperature QAHE from in-plane magnetization in atomic crystal layers of group-V elements.
\end{abstract}

\pacs{
73.23.-b,  
03.65.Vf,  
73.43.-f,  
71.10.Pm   
}

\maketitle

\textit{Introduction---.} Recent years have witnessed both theoretical and experimental progresses towards the understanding and realization of quantum anomalous Hall effect (QAHE)~\cite{Haldane}, which exhibits insulating bulk and conducting chiral edge states that are topologically protected from backscattering and show potential applications in low-energy consumption electronics~\cite{Rev_Qiao, Rev_2D2, Rev_2D3, Rev_2D4, Rev_2D5}. On the experimental side, the QAHE has been observed in magnetically doped 3D topological-insulator thin films, however, under ultra-low temperatures~\cite{QAHE_MagTI_exp_XueQK_13, QAHE_MagTI_exp_Checkelsky_14, QAHE_MagTIexp_WangKL_14, QAHE_MagTI_exp_Moodera_15}. This has further stimulated the searching for high-temperature QAHE in topological thin films, e.g., by considering the charge-compensated $n$-$p$ codoping mechanism~\cite{Qiao_np}. On the theoretical side, in narrow or zero-gap systems, like graphene~\cite{QAHE_G_Qiao_10, QAHE_G_adatom_Qiao_11, QAHE_G_5d_Mokrousov_12,QAHE_G_AFM_Qiao_14}, silicene~\cite{QAHE_QSHE_QVHE_Si_Ezawa_13, QAHE_QVHE_Si_Yao_14}, and other atomic crystal layers~\cite{QAHE_Bi(111)_Mokrousov_12, QAHE_Bi_Mokrousov_13, Rev_Qiao}, QAHE has been extensively studied by applying out-of-plane magnetization, but however is not yet experimentally observed.

The realization of QAHE requires the breaking of time-reversal symmetry, which is achieved by out-of-plane magnetization in most QAHE systems. Alternatively, the in-plane one, which is more energetically preferred in thin magnetic substrates~\cite{PRL_MagThinFilm, Rev_MagThinFilm}, also exhibits capability to induce QAHE in atomic crystal layers that display broken out-of-plane mirror-reflection symmetry~\cite{yafei, liu_inplane}. Such kind of material candidates, however, are rather rare~\cite{yafei, liu_inplane, QAHE_InPlane_OsCl_16}. Recent experiment has reported the opening of tiny band gap around $10~$meV in Bi thin film~\cite{Exp_BiTI}, the structure of which satisfies the symmetry requirement of harboring QAHE from in-plane magnetization induced by proximity to ferromagnetic materials~\cite{Bi_XFJin1, Bi_XFJin2}. Its strong spin-orbit coupling and tiny bulk gap together suggest a possible large-gap QAHE from weak magnetization.

\begin{table}
  \centering
  \begin{tabular}{ccccccccc}
    \hline \hline
     $\epsilon_{s}$  &  $\epsilon_{px,py}$  & $\epsilon_{pz}$  & $V_{ss\sigma}$  & $V_{sp\pi}$  & $V_{pp\sigma}$  & $V_{pp\pi}$  & $V'_{pp\sigma}$  & $V'_{pp\pi}$   \\
    \hline
     $-8.90$ & $-0.39$ & $-0.41$ & $-0.65$ & 1.94 & 1.90 & $-0.55$ & 0.60 & $-0.20$ \\
    \hline \hline
  \end{tabular}
  \caption{Slater-Koster tight-binding parameters of Bi (111) bilayer with lattice constant $a=4.39~$\AA~and interlayer distance $d=1.73~$\AA~in the absence of spin-orbit coupling. Unit: eV. The unprimed and primed parameters correspond to nearest and next-nearest neighbor hoppings, separately. }\label{parameter}
\end{table}

\begin{figure*}
  \centering
  \includegraphics[width=18cm,angle=0]{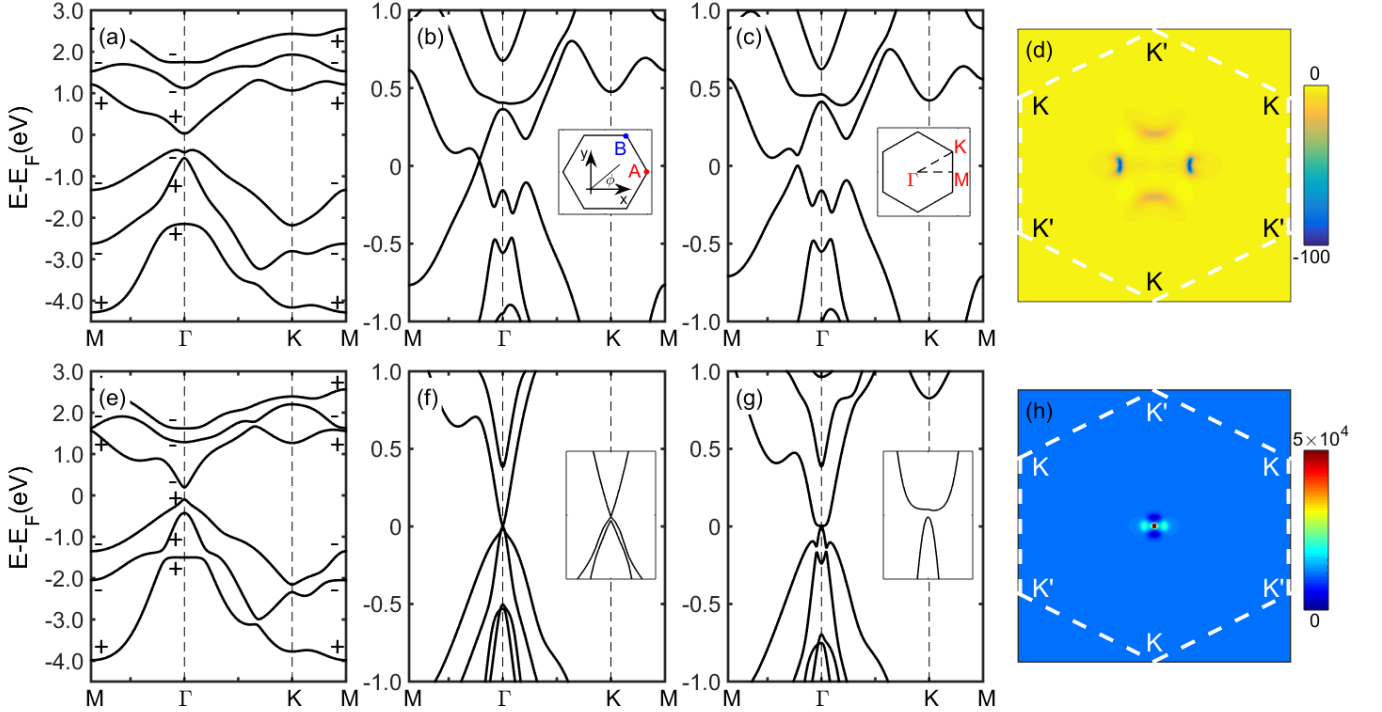}
  \caption{Band structure evolution for different magnetization strength $\lambda$ at spin-orbit coupling strengths of (a)-(c) $t_{\rm SO}=0.55~$eV and (e)-(g) $t_{\rm SO}=0.2~$eV. (a) and (e): Vanishing magnetization with $\lambda=0$. Even and odd parities of different eigenstates at time reversal invariant points ($\Gamma$ and $M$) are denoted by $+/-$. The parities of conduction and valence bands exchange at $\Gamma$ for these two cases suggesting a topological phase transition from (a) $\mathbb{Z}_2$ topological insulator to (e) topologically trivial band insulator. (b) and (f): Band structures at topological phase transition points with closing band gap when (b) $\lambda=0.64~$eV and (f) $\lambda=0.205~$eV. (c) and (g): Band structure with reopened band gap when (c) $\lambda=0.7~$eV and (g) $\lambda=0.36~$eV. (d) and (h): Berry curvature distributions for band structures in (c) and (g), which contribute to Chern number of $\mathcal{C}=-2$ and $+1$, separately. Inset in (b): Coordinate system and lattice structure with A (B) sublattice locating on the top (bottom) layer. Inset in (c): Schematic plot of first Brillouin zone and high symmetric lines along which we plot the band structure. Insets in (f) and (g): Zoom in of band structures near the band closing points. The magnetization is along $x$ direction with $\theta=\pi/2$ and $\phi=0$ in the above calculations. }\label{band}
\end{figure*}

In this Letter, we theoretically investigate the possibility of realizing QAHE from in-plane magnetization in atomic crystal layers of group-V elements with buckled honeycomb lattice based on the tight-binding model constructed by employing Slater-Koster two-center approximation. As concrete examples, we consider two limits of strong and weak spin-orbit couplings, which correspond respectively to $\mathbb{Z}_2$ topological insulator and topologically trivial band insulator. In the presence of in-plane magnetization, we find that QAHE can be realized in both cases to exhibit Chern numbers of $\mathcal{C}=\pm2$ and $\pm1$, separately. We then systematically study the phase diagram by manipulating the spin-orbit coupling strength, magnetization magnitude and its orientation. The former case exhibits a topologically nontrivial band gap that is farther exceeding the room temperature. Although this phase requires a stronger critical magnetization, it can be effectively reduced by applying strain. For the latter case, the critical magnetization depends highly on the spin-orbit coupling and can even reach close to zero. The resulting tiny topological band gap can be effectively enlarged by orders through slightly tilting the magnetization orientation from in-plane. Our findings not only provide a new class of materials that can harbor QAHE by breaking the time-reversal symmetry from in-plane magnetization, but also move forward to the possible experimental realization in 2D atomic crystal layers of group-V elements.

\begin{figure*}
	\centering
	\includegraphics[width=18cm,angle=0]{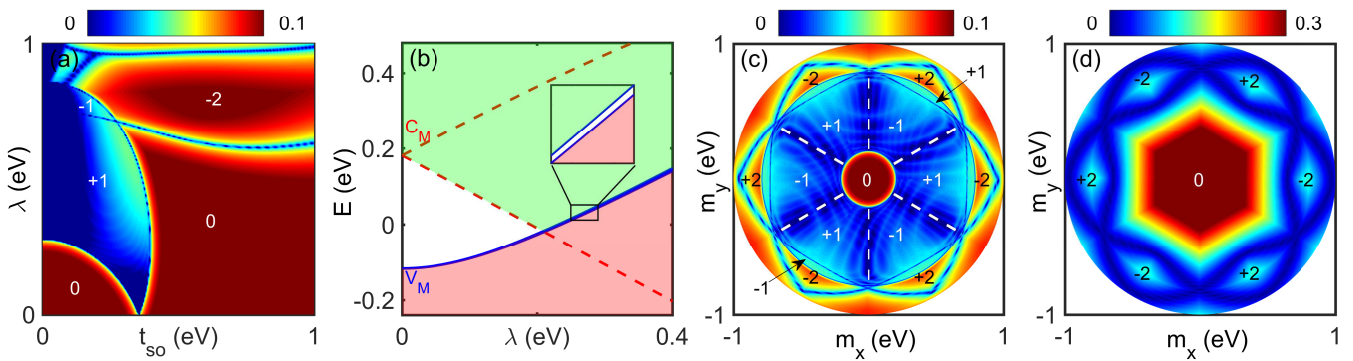}
	\caption{(a): Phase diagram \textit{v.s.} spin-orbit coupling strength $t_{\rm SO}$ and in-plane magnetization magnitude $\lambda$. The direct band gap is shown in color in unit of eV. Chern number of each topological phase is shown. (b): Evolution of energy levels at $\Gamma$ as function of magnetization $\lambda$. Areas covered with light green (red) is the unoccupied (occupied) energy regime with white regime being the direct band gap at $\Gamma$. Red dashed lines and blue solid lines: Energy levels split from conduction band minimal ($C_M$) and valence band maximal ($V_M$) at $\lambda=0$. (c) and (d): Phase diagram for different in-plane magnetization in $m_x$-$m_y$ plane for (c) $t_{\rm SO}=0.2~$eV and (d) $0.55~$eV.}\label{phi}
\end{figure*}

\textit{Model Hamiltonian---.} The tight-binding Hamiltonian is constructed by employing the two-center approximation on the orthogonal basis of \{$|s\rangle,|p_x\rangle,|p_y\rangle,|p_z\rangle$\}$\otimes$\{$|\uparrow\rangle, |\downarrow\rangle$\}~\cite{liu, book}:
\begin{align}
\label{tHamiltonian}
H & =  \sum_{i\alpha} \epsilon_{\alpha} c_{i\alpha}^{\dag} c_{i\alpha}  -\sum_{i\alpha,j\beta}t_{ij,\alpha\beta}c_{i\alpha}^{\dag}c_{j\beta}\\
 \nonumber & +t_{\rm SO} \sum_i c_{i\alpha}^{\dag} \bm{l} \cdot \bm{s} c_{i,\beta} \nonumber + \lambda \sum_{i\alpha} c_{i\alpha}^{\dag} \bm{\hat{m}} \cdot  \bm{s} c_{i\alpha},
\end{align}
where $c_{i\alpha}^{\dag}=(c_{i\alpha \uparrow}^{\dag},c_{i\alpha \downarrow}^{\dag})$ is the creation operator of an electron at the \emph{i}-th atomic site with $\uparrow/\downarrow$ and $\alpha/\beta$ representing spin up/down and different orbitals, respectively. The first term is the on-site energy, which is degenerate for $p_x$ and $p_y$ orbitals due to the three-fold rotation symmetry while that of $p_z$ orbital is slightly different. The second term stands for the hopping energy up to the next-nearest neighbors with an amplitude of $t_{ij,\alpha\beta}$, which are functions of Slater-Koster parameters listed in Tab.~\ref{parameter}~\cite{book}. The third term represents the intra-atomic spin-orbit coupling with a strength of $t_{\rm SO}$ with ${\bm{s}}=(s_x, s_y, s_z)$ and ${\bm{l}}=(l_x, l_y, l_z)$ being respectively Pauli matrices and orbital-angular-momentum operators, and can be expressed as:
\begin{equation}
H_{\rm SO}= t_{\rm SO}
\begin{pmatrix}
0 & 0 & 0 & 0\\
0 & 0 & -is_z & is_y\\
0 & is_z & 0 & -is_x\\
0 & -is_y & is_x & 0
\end{pmatrix}.
\end{equation}
The last term corresponds to an exchange coupling between electron and magnetization that is defined as $(m_x,m_y,m_z)$$\doteq$$\lambda \bm{\hat{m}} $$=$$\lambda (\sin \theta \cos \phi, \sin \theta \sin \phi, \cos \theta)$, where $\lambda$ is the magnetization strength and $\theta$/$\phi$ are respectively polar/azimuthal angles. It is noteworthy that this tight-binding model is generally applicable to lattices with $s$- and $p$-orbitals, like the atomic crystal layers of group-IV and -V elements. Without loss of generality, we perform our numerical study by employing parameters of Bi (111) bilayer as listed in Tab.~\ref{parameter} that are extracted from first-principles calculations by using non-linear least-square fitting. The detailed lattice and electronic structures from first-principles calculations are included in Supplemental Materials. In addition, we also consider the effect of biaxial strain that enlarges the lattice constants while the interlayer distance is reduced according to the Poisson's ratio that is set to be $0.335$~\cite{PoissonRatio}. The Slater-Koster hopping parameters are assumed to exhibit inverse square dependence on the interatomic distance~\cite{book}.

\textit{Band Structure---.} One can obtain the band structure by diagonalizing the momentum-space tight-binding Hamiltonian. In the absence of magnetization, Figs.~\ref{band}(a) and \ref{band}(e) display respectively the band structures for $t_{\rm SO}=0.55$ and $0.20~$eV, where even and odd parities of wavefunctions at the time-reversal invariant momenta ($\Gamma$ and M points) are labelled as ``$+/-$". As a comparison, one can see that the parities of conduction and valence bands exchange at $\Gamma$ point, indicating a topological phase transition from $\mathbb{Z}_2$ topological insulator to topologically trivial band insulator~\cite{z2}, which is in consistent with previous works~\cite{QSHE_Bi_Murakami_06, QSHE_Bi_BandStr_Blugel_08}. The presence of magnetization breaks the time-reversal symmetry and thus the $\mathbb{Z}_2$ topological phase. Specifically, in below we apply the in-plane magnetization ($\theta=\pi/2$) along $x$-axis ($\phi=0$), which meets the symmetry criteria by breaking all the mirror reflection symmetries of the buckled honeycomb lattice~\cite{yafei, liu_inplane}.

For strong spin-orbit coupling, we find that the increase of $\lambda$ shrinks the band gap that closes at $\lambda=0.64~$eV as indicated in Fig.~\ref{band}(b), where a Dirac cone appears on the high-symmetric line $M$-$\Gamma$. It is noteworthy that another Dirac cone also appears (not shown) due to the inversion symmetry of this system~\cite{yafei}. Further increase of $\lambda$ reopens the band gap inducing a QAHE with Chern number of $\mathcal{C}=-2$ obtained by integrating the Berry curvature over the first Brillouin zone according to the formula~\cite{Rev_Berry}:
\begin{equation}\label{Chern}
\mathcal{C}=\frac{1}{2\pi}\sum_{n}\int_{\rm BZ}d^{2}k\Omega_{n}(\bm{k}),
\end{equation}
where $\Omega_n(\bm{k})$ is the Berry curvature at momentum $\bm{k}$ of the $n$-th band~\cite{Rev_Berry, QAHE_G_Qiao_10}. The summation of $n$ runs over all occupied bands below the Fermi energy. In Fig.~\ref{band}(d), we plot the Berry curvature $\Omega({\bm{k}})$ obtained by summing over all occupied bands, where two negative peaks appear at the high-symmetric line $M$-$\Gamma$-$M$. The Berry curvature near each peak contributes to a Chern number of $-1$, resulting in a total Chern number of $\mathcal{C}=-2$.

The realization of QAHE does not necessarily require the host material system to be $\mathbb{Z}_2$ topological insulator before introducing ferromagnetism. For weak spin-orbit coupling case where the bands are topologically trivial [see Fig.~\ref{band}(e)], the presence of in-plane magnetization can also close the band gap as shown in Fig.~\ref{band}(f) where a Dirac cone appears at $\Gamma$ point. By further increasing $\lambda$, the system also enters a QAHE phase [see Fig.~\ref{band}(g)] but the resulting Chern number is $\mathcal{C}=1$. And the corresponding Berry curvature becomes centered around $\Gamma$ point with a high peak as plotted in Fig.~\ref{band}(h). In both cases, the Berry curvature distributes symmetrically about $\Gamma$ point guaranteed by the inversion symmetry invariance.

\textit{Effects of Spin-Orbit Coupling and In-Plane Magnetization Orientation---.} The influence of spin-orbit coupling strength $t_{\rm SO}$ on Chern number is systematically explored as displayed in Fig.~\ref{phi}(a) where the phase diagram is shown in the ($t_{\rm SO}$, $\lambda$) plane, with the color representing the direct band gap. At vanishing magnetization, i.e., $\lambda=0$, one can see that the band gap first closes and then reopens with the increase of $t_{\rm SO}$, indicating a topological phase transition from a trivial band insulator to a $\mathbb{Z}_2$ topological insulator at the critical point $t_{\rm SO}^{\rm c} \simeq 0.36~$eV. 
For $t_{\rm SO} > t_{\rm SO}^{\rm c}$, the presence of magnetization generally induces $\mathcal{C}=-2$ QAHE at relatively large magnetization strength $\lambda$. In this regime, one can see that the direct QAHE gap can be very large. Although the indirect band gap might be smaller, it is still in the order of $0.1~$eV that is farther exceeding the energy scale of room temperature and thus provides the possibility of realizing high-temperature QAHE. 

For $t_{\rm SO} \lesssim t_{\rm SO}^{\rm c}$, the introduction of in-plane magnetization can lead to the formation of a $\mathcal{C}=1$ QAHE phase. The critical magnetization strength $\lambda_c$ shows strong dependence on $t_{\rm SO}$ and decreases to zero as it approaches $t_{\rm SO}^{\rm c}$. As the magnetization strength increases, one can find various topological phase transitions to QAHE with Chern numbers $\mathcal{C}=-1$, $2$, and $0$ in sequence. Comparing to the $\mathcal{C}=1$ phase, these phases have smaller phase space and require larger magnetization strength. Thus we focus on the first phase in below. Despite the spin-orbit coupling strength in the order of $0.1~$eV, one ]can find that the band gap of $\mathcal{C} = 1$ QAHE is rather small. This can be understood from the band structure evolution when the magnetization strength $\lambda$ increases as shown in Fig.~\ref{phi}(b), where the occupied (unoccupied) energy regime is filled with light red (green). One can find that, the increase of $\lambda$ induces sizable energy splitting of the conduction band minimal ($C_M$) as plotted by red dashed lines while the splitting of valence band maximal ($V_M$) energy level is extremely small as shown in the inset of Fig.~\ref{phi}(b). After the band intersection between red dashed and blue solid lines, the two states represented by blue lines become separately conduction band minimal and valence band maximum with an extremely small band gap [See more details in Supplementary Material].

The resulting QAHE is robust when the magnetization changes in $m_x$-$m_y$ plane as shown in Figs.~\ref{phi}(c) and \ref{phi}(d) for different spin-orbit couplings $t_{\rm SO}=0.2$ and $0.55~$eV, separately. One can find that both phase diagrams show three-fold rotation symmetry since the magnetization orientations along $\phi + 2n\pi/3$ with integer $n$ are expected to be equivalent and give rise to the same topological phases due to the three-fold rotation symmetry of pristine Bi (111) bilayer. When the magnetization is along $\phi=\pi/6+ 2n\pi/3$ that preserves mirror reflection symmetry of Bi, no QAHE appears and the band gap is either vanishing or topologically trivial. This is similar to that of silicene and agrees with the symmetry analysis discussed in Ref.~\onlinecite{yafei}. Nevertheless, comparing to silicene~\cite{yafei}, the critical magnetizations required to reach QAHE for both $t_{\rm SO}=0.2$ and $0.55~$eV are much smaller. One can further manipulate the critical magnetization and band gap by changing spin-orbit coupling strength via alloy with Sb and P~\cite{QAHE_Bi(111)_Mokrousov_12} or by external means of tilting or strain as shown in below. 

\begin{figure}
  \centering
  \includegraphics[width=8cm,angle=0]{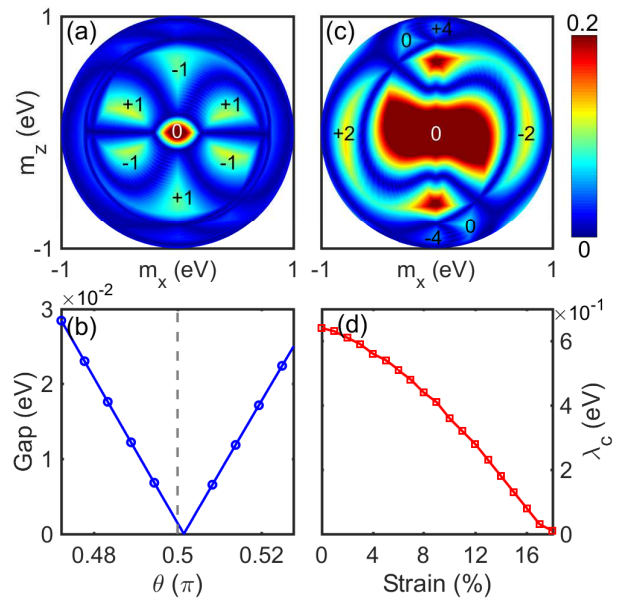}
  \caption{(a) and (c): Phase diagram \textit{v.s.} $m_x$-$m_z$ with spin-orbit coupling of  (a) $t_{\rm SO}=0.2~$eV and (c) $0.55~$eV. Direct band gap is shown in color in unit of eV. (b): Direct band gap \textit{v.s.} polar angle $\theta$ with $\lambda=0.36~$eV. (d): Direct band gap \textit{v.s.} biaxial strain for $t_{\rm SO}=0.55~$eV with magnetization pointing along $x$ axis.}\label{inout}
\end{figure}

\textit{External Manipulations---.} Although the QAHE phase with $\mathcal{C}=\pm1$ exhibits tiny bulk band gap, it can be effectively increased by tilting the magnetization from in-plane to introduce $z$ component. As shown in Fig.~\ref{inout}(a), we plot the direct band gap for different magnetizations in $m_x$-$m_z$ plane at fixed spin-orbit coupling $t_{\rm SO}=0.2~$eV. One can find that the tilting of magnetization from $x$ direction to either $+z$ or $-z$ can dramatically enlarge the band gap but with different Chern numbers. Our calculation shows that, by changing the magnetization from $x$-axis ($\theta=\pi/2$) to $z$-axis, the gap is increased by one order for a variation of $\theta$ by only $0.02\pi$ as plotted in Fig.~\ref{inout}(b). When the magnetization is rotated to $-z$ direction, however, the band gap first closes and then reopens inducing a phase transition to QAHE of $\mathcal{C}=-1$. The critical angle is around $(0.5+0.0014)\pi$ and a band gap of about $20~$meV can be found through changing $\theta$ by only $0.02\pi$, similar to the $\mathcal{C}=1$ case. 

For strong spin-orbit coupling case, such sensitivity of topological phase to $\theta$ is absent as shown in Fig.~\ref{inout}(c). One can find that the tilting of magnetization orientation from in plane can also close the band gap and induce topological phase transition to either topologically trivial band insulator or QAHE with opposite Chern numbers. However, the critical angle is quite large, indicating the robustness of the QAHE from in-plane magnetization. We further note that the critical magnetization required to realize QAHE is larger than that for weak spin-orbit coupling case. Nevertheless, as shown in Fig.~\ref{inout}(d), the critical magnetization $\lambda_c$ can be effectively reduced even close to zero by applying strain, which inevitably exists when Bi is placed on top of certain substrates~\cite{QSHE_Bi_Bansil_13}. 

\textit{Summary and Discussion---.} In summary, by applying in-plane magnetization, we theoretically confirmed the presence of QAHE in atomic crystal layers of group-V elements with buckled honeycomb lattice based on the $sp^3$ Slater-Koster tight-binding model. We numerically find that, starting from topologically trivial band insulator, QAHE of Chern numbers $\mathcal{C}=\pm1$ is formed with a small band gap, which can be effectively enlarged by orders via slightly tilting the magnetization from in-plane. For topologically nontrivial insulator, the Chern numbers are $\pm2$ with a large band gap that is farther exceeding the room-temperature energy scale. Although the critical magnetization strength is also larger, it can be effectively reduced by applying strain. Abundant topological phases appears when the magnetization orientation varies in $x$-$z$ plane.

Although our numerical calculations are performed based on the orbital parameters of Bi, the model is general and can capture the qualitative properties of the electronic structures of Sb, As and P with buckled structures. Therefore, our conclusions can directly generalize atomic layers of group V elements, making them potential candidates to harbor QAHE from in-plane magnetization. Moreover, for these structures, it has been reported that strain and electric field can effectively reduce their band gap~\cite{QSHE_Bi_Bansil_13, QSHE_Sb_Wang_13, QSHE_As_Ezawa_15, QSHE_P_Zunger_15}. Thus weak magnetization is expected to induce QAHE in these systems. Apart from the buckled bilayer systems, these elements also exhibit stable puckered structures~\cite{Sb_As} or other lattice structures~\cite{NewStr_GroupV}. These structures as well as their multilayer counterparts all satisfy the symmetry criteria of harboring QAHE from in-plane magnetization. Especially, the multilayers and some structures are reported to show smaller band gaps, indicating weaker critical magnetizations~\cite{QSHE_Bi_Liu_11, QSHE_Bi_Wu_11, Bi_Gap_ML_16}. A very recent experiment suggests a band gap of around $10~$meV for Bi thin film~\cite{Exp_BiTI}, which makes it possible to induce QAHE in Bi multilayer with a very weak magnetization. Further exploration on inducing in-plane ferromagnetism from adatoms or substrates from first-principles calculations is still desired to provide more detailed suggestions for experimental realizations.

\textit{Acknowledgements---.} We thank the financial support from National Key R \& D Program (2016YFA0301700), NNSFC (11474265), and the China Government Youth 1000-Plan Talent Program. L. Z. is supported by Innovation Program of Guangdong (608225320279) and Shenzhen Peacock Program (KQTD2016022619565991). The Supercomputing Center of USTC is gratefully acknowledged for the high-performance computing assistance.

$^\ddagger$ P. Zhong and Y. Ren contributed equally to this work.

\end{document}